%% file: GWMC-conference-format.tex
\documentclass{vgtc}          
\usepackage{amsmath}
\usepackage{amssymb}
\usepackage{amsthm}
\usepackage{multirow}

\usepackage{microtype}                 
\PassOptionsToPackage{warn}{textcomp}  
\usepackage{textcomp}                  
\usepackage{mathptmx}                  
\usepackage{times}                     
\usepackage{cite}                      
\usepackage{tabu}                      
\usepackage{booktabs}  

\newcommand {\mm}[1]        {\ifmmode{#1}\else{\mbox{\(#1\)}}\fi}

\newcommand{\Rspace}        {\mm{\mathbb{R}}}
\newcommand{\Mspace}        {\mm{\mathbb{M}}}

\newcommand{\Ccal}        {\mm{\mathcal{C}}}

\newcommand{\SN}        {\mm{\mathsf{Sinusoidal}}}
\newcommand{\WD}        {\mm{\mathsf{Wind}}}
\newcommand{\NS}        {\mm{\mathsf{NavierStokes}}}
\newcommand{\RS}        {\mm{\mathsf{RedSea}}}
\newcommand{\HC}        {\mm{\mathsf{HeatedCylinder}}}

\newcommand{\SG}       {\mm{\mathsf{RotatingGaussian}}}
\newcommand{\CG}       {\mm{\mathsf{RandomGaussian}}}

\newcommand{\para}[1]  {\vspace{1mm}\noindent{\textbf{#1}}}

\newcommand{\onevec}[1]     {\mm{\mathbf{1}_{#1}}}

\graphicspath{{./figs/}{./figsdraft/}}

\onlineid{1137}

\vgtccategory{Research}

\title{Comparing Morse Complexes Using Optimal Transport:\\An Experimental Study}

\author{Mingzhe Li\thanks{e-mail: mingzhe.li@utah.edu}\\ 
        \scriptsize University of Utah 
\and Carson Storm\thanks{e-mail: u1120712@utah.edu}\\ 
     \scriptsize University of Utah 
\and Austin Yang Li\thanks{e-mail: u1364758@utah.edu}\\ 
     \scriptsize University of Utah 
\and Tom Needham \thanks{e-mail: tneedham@fsu.edu}\\ 
     \scriptsize Florida State University 
\and Bei Wang \thanks{e-mail: beiwang@sci.utah.edu}\\ 
     \scriptsize University of Utah}
     
\abstract{
    \input{sec-abstract.tex}

}

\keywords{Morse Complexes, topological data analysis, optimal transport, topology in visualization}

\CCScatlist{
  \CCScatTwelve{Human-centered computing}{Visualization}
}

\begin{document}

\firstsection{Introduction}
\maketitle

\input{sec-introduction}

\input{sec-related-work}

\input{sec-background}

\input{sec-method}

\input{sec-results}

\input{sec-classify}

\input{sec-conclusion}

\acknowledgments{
This project was partially funded by NSF IIS-2145499, NSF IIS-1910733, NSF DMS-2107808, and DOE DE-SC0021015. 
}

\bibliographystyle{abbrv}
\bibliography{GWMC-conference-format.bib}


\end{document}


\maketitle









\appendix
\input{sec-ot-example}

\input{sec-datasets}

\input{sec-parameters}

\input{sec-runtime}

\input{sec-temporal-analysis}
\input{sec-partial-definition}
\input{sec-metric-property}

\bibliographystyle{abbrv}
\bibliography{GWMC-refs.bib}

%% file: sec-abstract.tex
Morse complexes and Morse-Smale complexes are topological descriptors popular in topology-based visualization.
Comparing these complexes plays an important role in their applications in feature correspondences, feature tracking, symmetry detection, and uncertainty visualization. 
Leveraging recent advances in optimal transport, we apply a class of optimal transport distances to the comparative analysis of Morse complexes.  
Contrasting with existing comparative measures, such distances are easy and efficient to compute, and naturally provide structural matching between Morse complexes.  
We perform an experimental study involving scientific simulation datasets and discuss the effectiveness of these distances as comparative measures for Morse complexes. 
We also provide an initial guideline for choosing the optimal transport distances under various data assumptions.

%% file: sec-introduction.tex
Morse complexes and Morse-Smale complexes are gradient-based topological descriptors of scalar fields. 
They have numerous applications in scientific visualization, such as feature correspondences~\cite{FengHuangJu2013}, feature tracking~\cite{KastenReininghausHotz2011, ReininghausKastenWeinkauf2012, NarayananThomasNatarajan2015, KuhnEngelkeFlatken2017,
     SchnorrHelmrichChilds2019, SchnorrHelmrichDenker2020, 
      RieckSadloLeitte2020}, symmetry detection~\cite{ThomasNatarajan2013}, structural change detection~\cite{NarayananThomasNatarajan2015}, and uncertainty visualization~\cite{AthawaleMaljovecYan2022}. 
Only a few comparative measures have been developed for these complexes and their variants~\cite{NarayananThomasNatarajan2015,FengHuangJu2013,SchnorrHelmrichDenker2020}. 

On the other hand, recent years have seen the successful application of optimal transport to graph analysis. 
Namely, the Wasserstein distance \cite{villani2009optimal}, Gromov-Wasserstein distance \cite{Memoli2007,Memoli2011}, and their  variants \cite{vayer2020fused,ChapelAlayaGasso2020} have been used extensively for graph matching and comparison. 

Leveraging recent advances in optimal transport (OT), we apply, \emph{for the first time}, a class of OT-type distances to the comparative analysis of Morse complexes. 
Contrasting with existing comparative measures, such distances are easy and efficient to compute, and provide explicit structural matching   between Morse complexes. 
Our main contribution is to provide experimental studies to evaluate the effectiveness of OT-type distances in terms of feature correspondences and classification, and to provide an initial guideline for choosing the OT-type distances under various data assumptions.

%% file: sec-related-work.tex
\section{Related Work}
\label{sec:related-work}

\para{Optimal transport for graph matching and comparison.} 
M\'{e}moli first introduced Gromov-Wasserstein (GW) distances \cite{Memoli2007,Memoli2011} for the  comparison of metric measure spaces. 
Theoretical works that expanded the scope of the GW framework (e.g., \cite{chowdhury2019gromov}) and new approaches in optimization \cite{PeyreCuturiSolomon2016} have made GW distances a popular tool for studying unregistered graphs. 
We briefly survey some of the related work here. 

Applications of GW distances for graph analysis in a machine learning setting were explored by Xu et al.~\cite{XuLuoZha2019,XuLuoCarin2019,xu2022representing}, with a focus on scalability and novel uses of the probabilistic nature of the metric. 
A Riemannian structure on the space of graphs endowed with the GW distance was established in \cite{ChowdhuryNeedham2020}, with a view toward statistical analysis of graphs. 
Connections between GW distances and spectral graph theory were provided in \cite{chowdhury2021generalized}, inspired by earlier work of M\'{e}moli in the setting of Riemannian manifolds \cite{memoli2009spectral}. 
Several variants of GW distances have been proposed~\cite{chowdhury2021quantized,vincent2022semi,sejourne2021unbalanced}, frequently motivated by applications in graph analysis, such as the fused GW~\cite{VayerCourtyTavenard2019,vayer2020fused} and the partial GW distances~\cite{ChapelAlayaGasso2020}; these are discussed in detail below (see \cref{sec:background}).
Finally, the GW framework has been successfully applied to compare certain topological descriptors. In particular, it has been used as a tool for summarizing merge tree ensembles~\cite{li2021sketching}, feature tracking~\cite{LiYanYan2023}, and comparing merge trees and Reeb graphs endowed with additional topological attributes~\cite{curry2022decorated,curry2023topologically}. 

\para{Comparative measures for Morse and Morse-Smale complexes} have previously focused on comparing the graphs derived from the complexes; see~\cite{YanMasoodSridharamurthy2021} for a survey.  
Feng et al.~\cite{FengHuangJu2013} studied \emph{feature graphs}, which are 1D skeletons of simplified Morse-Smale complexes, to represent non-rigidly deformed surfaces. 
They used a minimum-cost matching algorithm to compare feature graphs. 
Thomas and Natarajan~\cite{ThomasNatarajan2013} used \emph{augmented extremum graphs} to detect symmetry in scalar fields. 
They used the geodesic distances between extrema and earth mover's distance between histograms of seed regions.
Narayanan et al.~\cite{NarayananThomasNatarajan2015} defined a distance between extremum graphs by forming a complete graph between all pairs of extrema and computing the maximum distortion of the node sets and the edge sets.  
To the best of our knowledge, this paper presents the first time OT-type distances are used as comparative measures for Morse complexes.

%% file: sec-background.tex
\section{Background}
\label{sec:background}
We review Morse complexes and their 1D skeletons, referred to as Morse graphs.
We also introduce notions of optimal transport (OT) type distances that are applicable in our comparative analysis.   

\para{Morse complexes and Morse graphs.}
We focus on 2D Morse complexes in this paper.  
Let $f: \Mspace \to \Rspace$ be a Morse function defined on a 2D manifold with boundary $\Mspace \subset \Rspace^2$, with gradient denoted $\nabla f$.  
A point $x \in \Mspace$ is a \textit{critical point} if $\nabla f=0$;  otherwise, it is a \textit{regular point}.  
There are three types of critical points: local maxima, local minima, and saddles. 
The \emph{integral line} of a regular point is the maximal path through the point whose tangent vectors align with the gradient. 
The \textit{descending manifold} surrounding a critical point contains the point itself and all regular points whose integral lines end at the critical point. 
Critical points (local minima and saddles) are the 0-cells, integral lines connecting these critical points are the 1-cells, and the rest of the domain makes up the 2-cells of a complex called the \emph{Morse complex} of $f$. Its 1D skeleton consisting of 0- and 1-cells is referred to as the \emph{Morse graph}.

For example, the 3D graph of a 2D function $f$ is shown in~\cref{fig:synthetic} (bottom left), where its domain is segmented into nine descending manifolds (surrounding nine local maxima). 
Its corresponding Morse graph is embedded in the graph of $f$ in cyan; it is also shown in 2D from a top-down viewpoint (top left).

\para{Measure networks.}
In our context, a Morse graph is modeled as a \emph{measure network}~\cite{chowdhury2019gromov}. That is, it is represented as a triple $G = (V, p, W)$, where $V$ is a finite set of nodes sampled from the Morse graph, $p: V \to [0, 1]$ is a probability density on $V$ (i.e., $p(x) \geq 0$ for all $x$ and $\sum_{x \in V} p(x) = 1$), and $W: V \times V \to \Rspace$ is a \emph{network function} that captures the relations between pairs of nodes; the particular choice of node density and network function used to represent a Morse graph is explained in \cref{sec:method}. Given a function $F:V \to A$, for some metric space $(A,d_A)$, we may also consider a Morse graph as an \emph{$A$-attributed measure network}, consisting of the data $G = (V,p,W,F)$. In this paper, we take our attribute space to be $\Rspace^2$ endowed with a Euclidean distance, and the attribute function $F$ assigns a node its location in the domain $\Mspace$. 

Given a pair of Morse graphs $G_1=(V_1, p_1, W_1, F_1)$ and $G_2=(V_2,  p_2, W_2, F_2)$, such that the codomain of $F_1$ and $F_2$ is a common metric space $(A,d_A)$,  we introduce some notation for the sake of simplicity. 
Let $n_1=|V_1|$ and $n_2=|V_2|$ and enumerate the node sets as $\{x_i\}_{i=1}^{n_1}$ and $\{y_j\}_{j=1}^{n_2}$, respectively. 
We write $W_1(i,k) := W_1(x_i,x_k)$, $p_1(i) := p_1(x_i)$, and $a_i := F_1(x_i)$. 
It is convenient to consider $W_1$ as a matrix in $\Rspace^{n_1 \times n_1}$ and $p_1$ as a vector in $\Rspace^{n_1}$. We use similar notation for $G_2$, except we write $b_j := F_2(y_j)$. 

We review a number of OT-type distances. They have the common feature of being defined in terms of measure couplings. A \emph{coupling} $C$ between probability measures $p_1$ and $p_2$ is a non-negative real-valued matrix representing a joint probability measure on $V_1 \times V_2$ whose row and column marginals agree with $p_1$ and $p_2$, respectively. 
The set of all couplings between $p_1$ and $p_2$ is denoted as 
\begin{align}
\label{eq:coupling}
    \Ccal = \Ccal(p_1, p_2) := \{C \in \Rspace_{+}^{n_1\times n_2} \mid C \onevec{n_2} = p_1, C^T \onevec{n_1} = p_2\},
\end{align} 
where $\onevec{n} = [1, 1, ..., 1]^T \in \Rspace^n$. 

The OT-type distances under consideration are described below.  Each OT-type distance is defined by an optimization problem over $\Ccal$. Intuitively, we think of an element of $\Ccal$ as a probabilistic ``soft" registration of the vertices of the Morse graphs. The optimization problems favor registrations which preserve intrinsic (node relations) and/or extrinsic (attribute) features as much as possible. Each distance depends on a parameter $q \in [1,\infty)$; the definitions can be extended to $q=\infty$, but we will generally fix $q = 2$ in experiments.

\para{Wasserstein distance \cite{villani2009optimal}.} The \emph{$q$-Wasserstein distance} between $G_1$ and $G_2$ is defined by  
\begin{align}
d^{W}_q(G_{1},G_{2})^q = \min_{C\in\Ccal} \sum_{i,j} d_A(a_i,b_j)^q C_{i,j}.
\label{eq:W}
\end{align}
This distance compares nodes based on attributes, but is agnostic to the network structure encoded by the $W$-functions.

\para{Gromov-Wasserstein distance \cite{Memoli2007}.}
The \emph{$q$-Gromov-Wasserstein distance} (GW) is defined as 
\begin{align}
d_{q}^{GW}(G_1, G_2)^q =  \min_{C\in\Ccal}\sum_{i, j, k, l}|W_1(i, k) - W_2(j, l)|^{q} C_{i,j}C_{k,l}
\label{eq:GW}
\end{align}
This distance compares the network structures of the Morse graphs encoded by the $W$-functions, but has no dependency on attributes.

\para{Fused Gromov-Wasserstein distance \cite{VayerCourtyTavenard2019}.}
With a trade-off parameter $\alpha \in [0,1]$, the \emph{$q$-Fused Gromov-Wasserstein distance} (FGW) between attributed Morse graphs $G_{1}$ and $G_{2}$ is defined as  
\begin{align}
& d^{FGW}_{q,\alpha}(G_{1},G_{2})^q =  \min_{C\in\Ccal}  \sum_{ i,j,k,l} [(1-\alpha) d_A(a_i,b_j)^q \nonumber \\
& \qquad \qquad \qquad \qquad  + \alpha |W_1(i,k) - W_{2}(j,l))|^q] C_{i,j}C_{k,l}.
\label{eq:fgw}
\end{align}
We have $d^{FGW}_{q,0} = d^W_q$ and $d^{FGW}_{q,1} = d^{GW}_q$, whereas $\alpha \in (0,1)$ yields a distance which depends on both network and attribute structures.
See the supplementary materials for a simple example illustrating the differences between these distances.

The distances defined above can be sensitive to outliers, since they are forced to match the full masses of $p_1$ and $p_2$. These constructions can be made more robust by using \emph{relaxed couplings}. That is, given $m \in [0,1]$, we may consider measures on $V_1 \times V_2$ with relaxed coupling constraints and a total mass given by $m$; that is, we define
\begin{align}
\label{eq:pw-couplings}
&\Ccal_m = \Ccal_m(p_1, p_2) \nonumber \\
&:= \{C \in \Rspace_{+}^{n_1 \times n_2} \mid C\onevec{n_{2}} \leq p_{1}, C^{T}\onevec{n_{1}} \leq p_{2}, \mathbf{1}_{n_{1}}^T C \onevec{n_{2}} = m \}. 
\end{align}  
Each of the OT-type distances $d^{W}_q$, $d^{GW}_q$ and $d^{FGW}_{q,\alpha}$ has an associated ``partial'' version, which depends on an additional mass parameter $m$. These are obtained by replacing optimization over $\Ccal$ in the above definitions with optimization over $\Ccal_m$. Intuitively, elements of $\Ccal_m$ still represent soft registrations between nodes, but with extra flexibility due to their ability to ignore some proportion of mass.
Respectively, the partial versions of the OT-type distances are denoted to as $d^{pW}_{q,m}$ for \emph{partial Wasserstein distance} (pW) \cite{ChapelAlayaGasso2020}, $d^{pGW}_{q,m}$ for \emph{partial Gromov-Wasserstein distance} (pGW) \cite{ChapelAlayaGasso2020} and $d^{pFGW}_{q,\alpha,m}$ for \emph{partial Fused Gromov-Wasserstein distance} (pFGW) \cite{LiYanYan2023}.
See the supplementary materials for formal definitions of the partial OT-type distances and a discussion of metric properties of all distances.

%% file: sec-method.tex
\section{Method}
\label{sec:method}
To compare Morse graphs using the OT-type distances discussed in \cref{sec:background}, we first model them as measure networks $G = (V, p, W)$. 
To emphasize well-connected nodes essential to the Morse graphs, we chose a $p$ measure based on the degrees of the nodes following~\cite{XuLuoZha2019}, in which
$
p(v) = \frac{\deg(v)}{\sum_i \deg{v_i}};
$ 
this heuristic choice gives extra importance to more highly connected nodes when searching for an optimal registration.
To capture the intrinsic pairwise relations between nodes in the Morse graph, we define $W(v, w)$ to be the shortest geodesic path length between $v$ and $w$ in $V$, following~\cite{LiYanYan2023}. 
Finally, to model a Morse graph as an $A$-attributed measure network, the attribute function assigns a node $v$ its location $(v_x, v_y)$ in the domain.  

To examine the structural alignment between Morse graphs, we apply a color transfer based on the coupling matrix between a source (reference) and a target Morse graph. 
We first assign a color to each node in the source using a gradient colormap based on node position. For each of the distances described above, we compute a minimizing coupling $C$. To each node $y_j$ in the target, we assign the color corresponding to the node $\mathrm{argmax}_{x_i} C(i,j)$ in the source. We then evaluate the quality of the alignment via visual inspection. 

\para{Implementation.} 
We compute the Morse graphs using TTK (1.1.0)~\cite{TiernyFavelierLevine2021} and ParaView (5.10.1)~\cite{Ahrens2005ParaViewAE}. 
The computation of the six OT-type distances is extended  from~\cite{LiYanYan2023}, which is available on GitHub~\cite{MCOpt}.   
Our implementation for the FGW (thus Wasserstein and GW) distances follows~\cite{vayer2020fused, VayerCourtyTavenard2019} and Python Optimal Transport (POT)~\cite{flamary2021pot}, which uses conditional gradient method for optimization. 
For partial OT-type distances, we add dummy nodes to carry over the probability mass to be ignored, following~\cite{ChapelAlayaGasso2020}.

%% file: sec-results.tex
\begin{figure*}[!ht]
    \centering
    \includegraphics[width=0.9\textwidth]{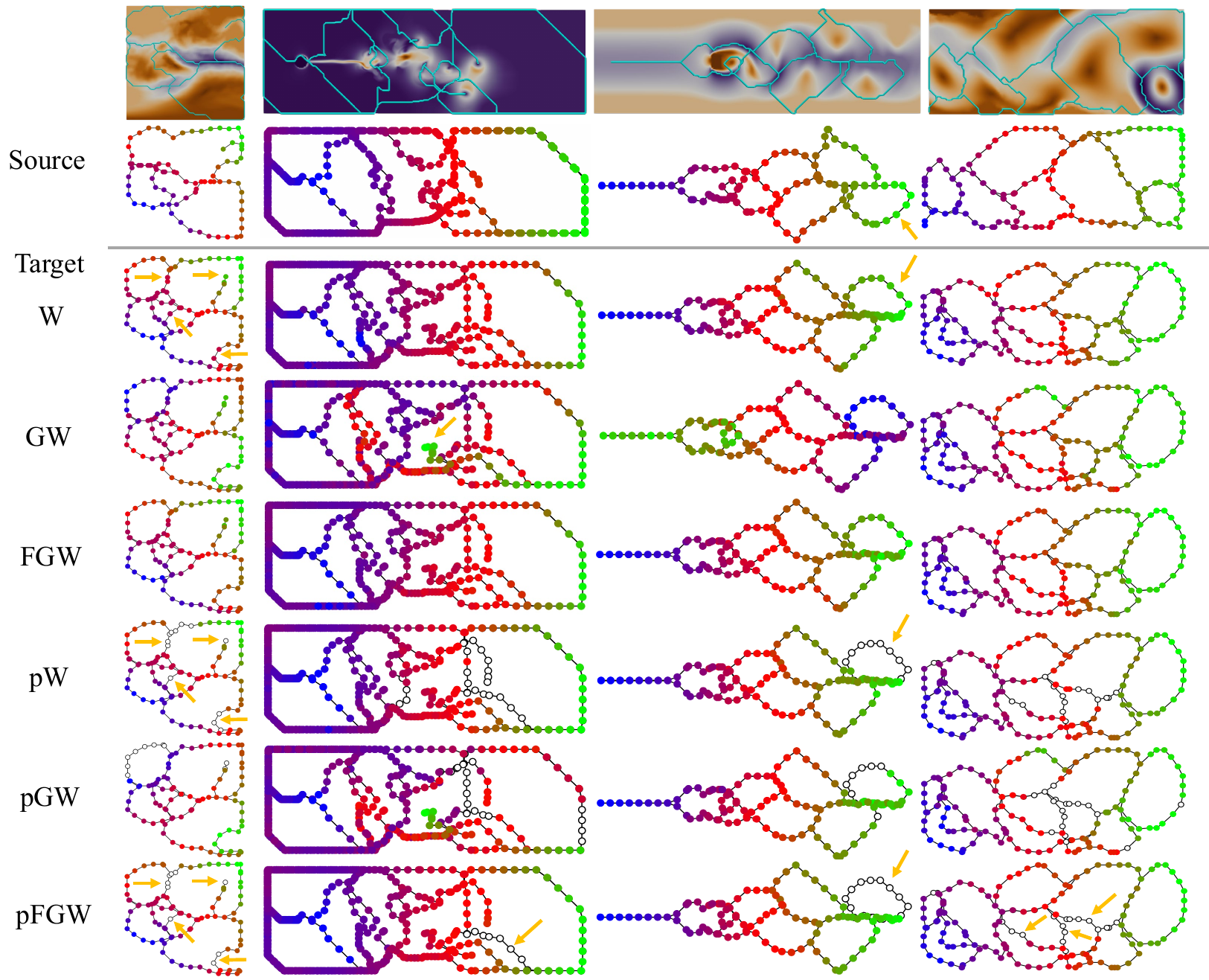}
    \vspace{-2mm}
    \caption{Structure alignments between the source and the target across all OT-type distances (from left to right) for the {\WD}, {\HC}, {\NS},  
    and {\RS} datasets, respectively. W: Wasserstein distance.}
    \label{fig:real-world}
    \vspace{-6mm}
\end{figure*} 

\section{Results}
\label{sec:results}

We study various OT-type distances using synthetic and real-world datasets.   
The first instance of each dataset is used as the source and shown in~\cref{fig:real-world} (1st and 2nd rows);
see supplementary materials for details on the datasets, parameter tuning, and runtime analysis. 

\subsection{An Overview With A Synthetic Dataset}
\label{sec:synthetic}

We first use the synthetic {\SN} dataset to compare the behaviors of six OT-type distances introduced in~\cref{sec:background}, where the target field is generated from the source field with added noise. 
We use a color transfer to highlight the structural alignment between the source and the target Morse graphs across all distances. 
\begin{figure}[h]
    \centering
    \includegraphics[width=0.9\columnwidth]{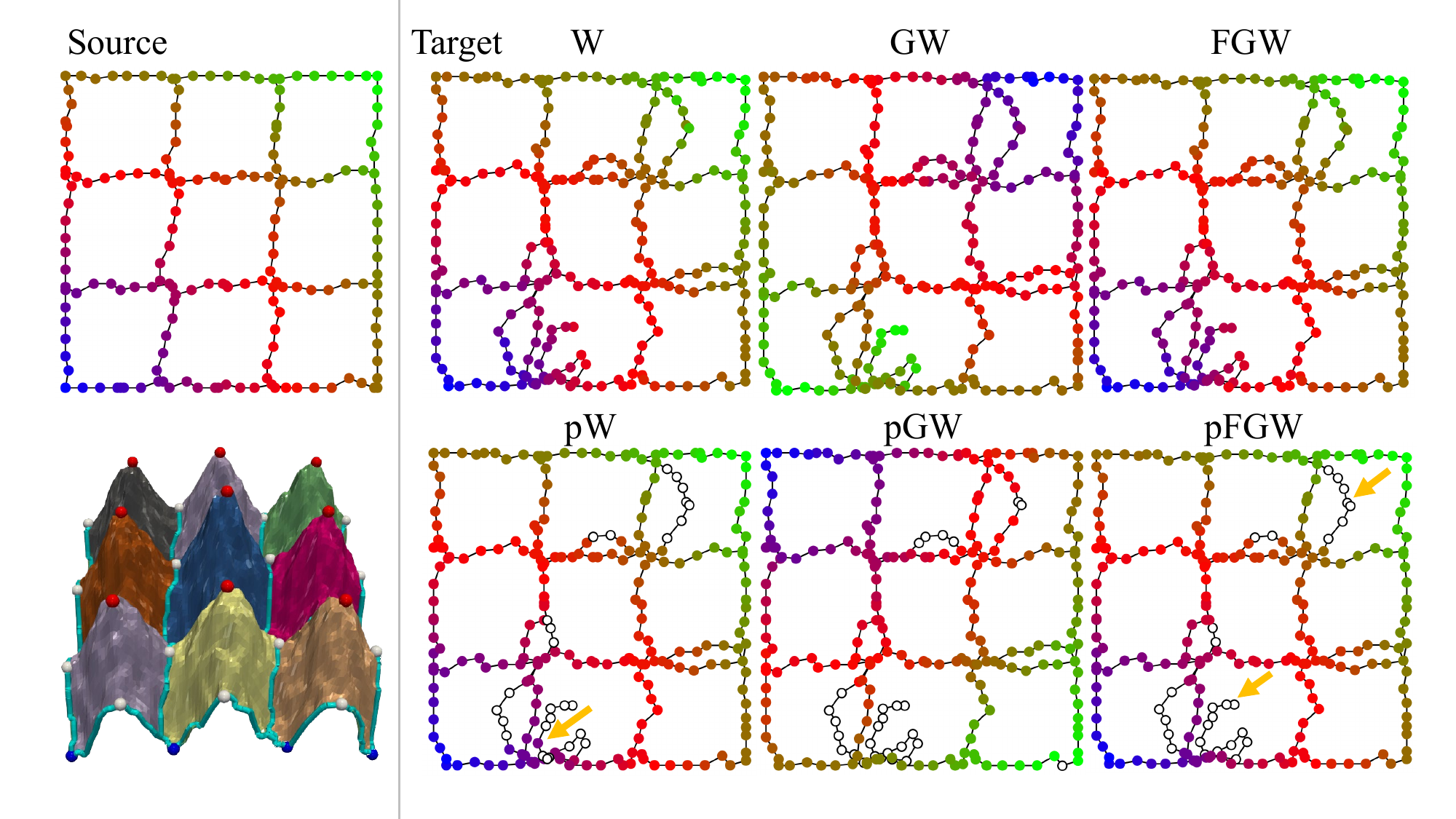}
    \vspace{-2mm}
    \caption{{\SN} dataset: structural alignments between the source and the target across all OT-type distances.}
    \label{fig:synthetic}
    \vspace{-2mm}
\end{figure}
As shown in~\cref{fig:synthetic} (top right), both the Wasserstein (W) and the FGW distance show reasonable structural alignments, even with the added noise in the target domain.  
However, using the GW distance, we observe that the bottom left of the source is aligned with the top right of the target. 
This is due to the fact that the GW distance focuses on the intrinsic relations among nodes in the Morse graph and is unaware of the symmetry within the dataset.  
Therefore, the GW distance may result in flipped or rotated alignments. 

On the other hand, the partial OT-type distances aim to be more robust by relaxing the couplings and thereby ignoring certain noisy features. 
As shown in~\cref{fig:synthetic} (bottom right), certain nodes in the target are hollow, which are ignored by partial OT during optimization. 
In particular, using the pFGW distance, a number of 1-cells in the target (indicated by orange arrows) that deviate from the source are ignored in the alignment, indicating its robustness against noise.  
 
Both pW and pGW distances are able to ignore subsets of the noisy features.
The result from pW is slightly worse than that of pFGW. For example, in the relaxed coupling of pW, a node (indicated by the orange arrow) becomes isolated because its neighboring nodes are ignored.
This is because the pW distance does not consider the intrinsic relations between nodes in the Morse graph. 
The pGW distance, on the other hand, results in an upside-down flipped alignment. 
Overall, for the {\SN} dataset, the pFGW distance performs the best for feature correspondences, followed by the pW distance.

\subsection{Real-World Datasets: Feature Correspondences}
\label{sec:real-world}

We now examine the OT-type distances using real-world datasets in~\cref{fig:real-world}. 
For a time-varying dataset, we use an earlier time step as  the source and a later time step as the target, that is, time steps $1$ and $10$ for {\WD}~\cite{NCEPData},  $800$ and $879$ for {\HC}~\cite{gerrisflowsolver}, and $60$ and $63$ for {\NS}~\cite{camarri05} datasets, respectively. 
For the ensemble dataset {\RS}~\cite{sciVis2020}, we compare ensemble members $1$ and $4$. 

\para{Wind dataset.} 
For full OT-type distances, the Wasserstein and the FGW distances align the global features reasonably well, resulting in a smooth color transfer between the source and the target.
The GW distance produces a top-down flipped alignment due to symmetry.   
Four noticeable structural differences exist between the source and target (indicated by orange arrows). 
Among the partial OT-type distances, pFGW distance performs the most robustly as it ignores these four differences by treating them as noisy features. The pW distance ignores all four features but also ignores two more nodes on the top boundary.
Again, pGW suffers from symmetry in the data. 
In summary, partial OT is useful when we are interested in capturing global similarities while ignoring local differences. 

\para{Heated Cylinder dataset.}
The differences between the source and the target are  minor. 
Both Wasserstein and FGW distances produce reasonable alignments, whereas the GW distance produces some noticeable misalignments in the center of the domain (indicated by an orange arrow). 
Using partial OT, the pFGW distance performs the best by ignoring the additional 1-cell in the target during alignment (indicated by orange arrows). 
In comparison, the pW and pGW distances perform less robustly than the pFGW distance.
For instance, some 1-cells are incorrectly ignored, also possibly due to inconsistent sampling density between the source and the target. 

\para{Navier Stokes dataset.}
Comparing the source and the target, we notice one additional loop on the top right corner of the target.
Both Wasserstein and FGW distances match such an additional loop in the target to a bottom-right loop in the source (indicated by orange arrows). 
This is undesirable but understandable, as these two distances are forced to match the full masses on the nodes between the source and the target. 
With partial OT, the pW and pFGW distances could ignore the additional loop in the coupling matrices. 
The pGW distance also ignores a part of the loop. 

\para{Red Sea dataset.}
The {\RS} dataset is an ensemble dataset. As a result, the target has a very different structure from the source.  
Nevertheless, we aim to align these two Morse graphs as much as possible. 
Results using the Wasserstein and the FGW distances  are similar. 
While it is trickier to tune the parameter for partial OT (in comparison with time-varying datasets), we observe that pW and pFGW distances ignore 1-cells at the center of the target to form bigger holes to better align with the source (that contains bigger holes and fewer loops).

%% file: sec-classify.tex
\subsection{Classification: Wasserstein vs. GW Distances}
\label{sec:classify}

We observe that the Wasserstein distance typically produces better alignments than the GW distance for the evolving scalar field datasets in   \cref{sec:real-world}. 
We now discuss potential scenarios when the GW distance outperforms the Wasserstein distance.

We introduce two synthetic datasets for classification tasks.
For the {\SG} dataset, we generate a mixture of two Gaussian functions with equal bandwidth and rotate the mixture at 100 evenly sampled angles (referred to as the \emph{Rotating Binary Gaussians}); similarly, we generate 100 rotations of a mixture of three Gaussian functions (referred to as the \emph{Rotating Trinary Gaussians}). 
For the {\CG} dataset, we generate 100 instances of randomly generated mixtures of two and three Gaussian functions with random centers and varying bandwidths, referred to as \emph{Random Binary Gaussians} and \emph{Random Trinary Gaussians},  respectively. 
We introduce random noises to increase the complexity of each dataset. 

\begin{figure}[!ht]
    \centering
    \includegraphics[width=0.9\columnwidth]{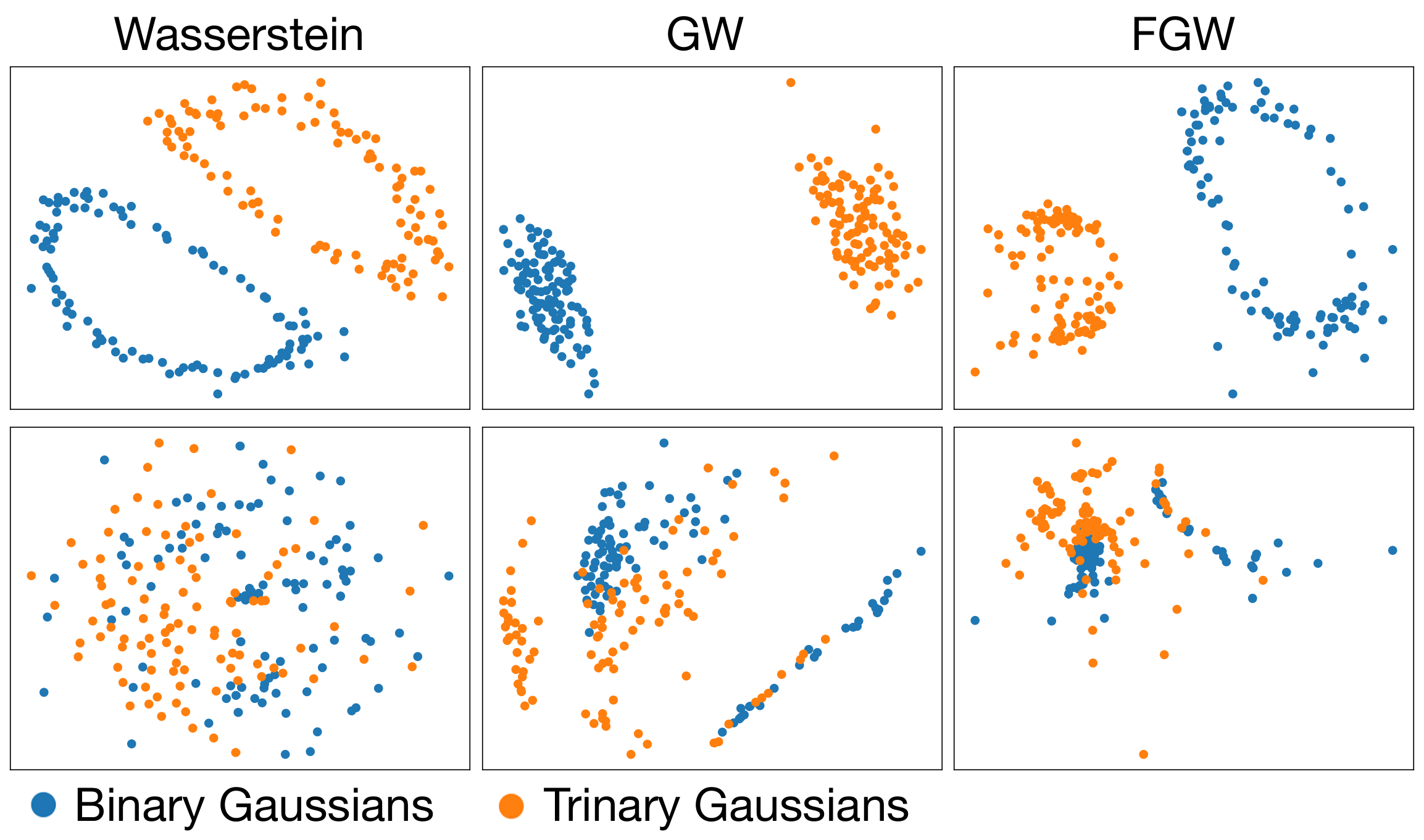}
    \vspace{-2mm}
    \caption{MDS projections for the {\SG} (top) and {\CG} (bottom) datasets.}
    \label{fig:gaussian-mds}
    \vspace{-6mm}
\end{figure}

We apply multi-dimensional scaling (MDS)~\cite{borg2005modern} to visualize the GW, FGW, and Wasserstein distances for each dataset, as shown~\cref{fig:gaussian-mds}. 
Blue points represent data instances from Binary Gaussians while orange points are from Trinary Gaussians. 
For the {\SG} dataset, all three distances produce visible clusters and clearly separate the two classes. 
The periodic behaviors of the dataset are clearly visible as loops from the MDS projection using the Wasserstein distance as it is sensitive to geometry. 
For the {\CG} dataset, we still observe some clustering effects using GW and FGW distances, but without clear separations between the classes. 

We then apply a k-Nearest Neighbors (kNN) classifier to explore the classification accuracies using these distances. 
For each dataset, we use an 80-20 split of training and testing data. 
We apply cross-validation to set $k=3$. We report the test accuracy and the F1-score for each class for all three distances, as shown in \cref{table:classification}. 
We observe better classification results using a GW distance in comparison with the Wasserstein distance. 
The property that GW distances are insensitive to the geometric locations and solely measure the difference between intrinsic graph structures becomes an advantage in this specific task.

\begin{table}[!ht]
\begin{center}
\vspace{-2mm}
\resizebox{1.0\columnwidth}{!}{
\begin{tabular}{|c|c|c|c|c|}
\hline
\textbf{Dataset}             & Distance    & Accuracy & \begin{tabular}[c]{@{}c@{}}F1-score \\ Binary Gaussians\end{tabular} & \begin{tabular}[c]{@{}c@{}}F1-score \\ Trinary Gaussians\end{tabular} \\ \hline
\multirow{3}{*}{{\SG}} & Wasserstein & 1.00     & 1.00                      & 1.00                       \\
                             & GW          & 1.00     & 1.00                      & 1.00                       \\
                             & FGW         & 1.00     & 1.00                      & 1.00                       \\ \hline
\multirow{3}{*}{{\CG}} & Wasserstein & 0.70     & 0.76                      & 0.60                       \\
                             & GW          & \textbf{0.85}     & \textbf{0.86}                      & \textbf{0.83}                       \\
                             & FGW         & 0.70     & 0.77                      & 0.57                       \\ \hline
\end{tabular}
}
\end{center}
\vspace{-6mm}
\caption{Test accuracy and F1-score for kNN classifiers ($k=3$) using the Wasserstein, GW, and FGW    distances.}
\label{table:classification}
\vspace{-6mm}
\end{table}

%% file: sec-conclusion.tex
\section{Guidelines and Future Work}
\label{sec:conclusion}

The Wasserstein distance optimizes the coupling by minimizing the Euclidean distances between matched points, whereas the GW distance solely preserves the intrinsic node relations. 
Combining these two components, the FGW distance preserves both Euclidean proximity and intrinsic similarity during alignment. 
Both the Wasserstein and FGW distances perform reasonably well in aligning the global structures between the source and the target. 
In comparison, the GW distance is less robust due to a lack of geometric information to ``anchor'' the aligned regions. 
If full structural alignment is required, then the FGW distance is typically the top choice for feature correspondence and comparison tasks, followed by the Wasserstein distance. 
If intrinsic graph structures are more important than extrinsic geometry, such as for certain classification tasks, then the GW distance outperforms the distances with a Wasserstein component. 

On the other hand, distances based on partial OT are generally more robust to noise and outliers than their full OT counterparts.
If partial structural alignment is allowed, the pFGW distance typically performs the best for feature correspondences, followed by the pW distance. Aside from being flexible, partial OT also helps highlight the structural differences between the source and the target.

Using time-varying datasets, these OT-type distances can be used in feature tracking. In particular, partial OT may be used to detect structural changes over time. Our comparative analysis could be easily extended to study Morse-Smale complexes and their variants such as the extremal graphs. This is left for future work.

%% file: sec-ot-example.tex
\section{Discussion on OT-type Distances}
\label{sec:ot-example}

\begin{figure}[!ht]
    \centering
    \includegraphics[width=0.98\columnwidth]{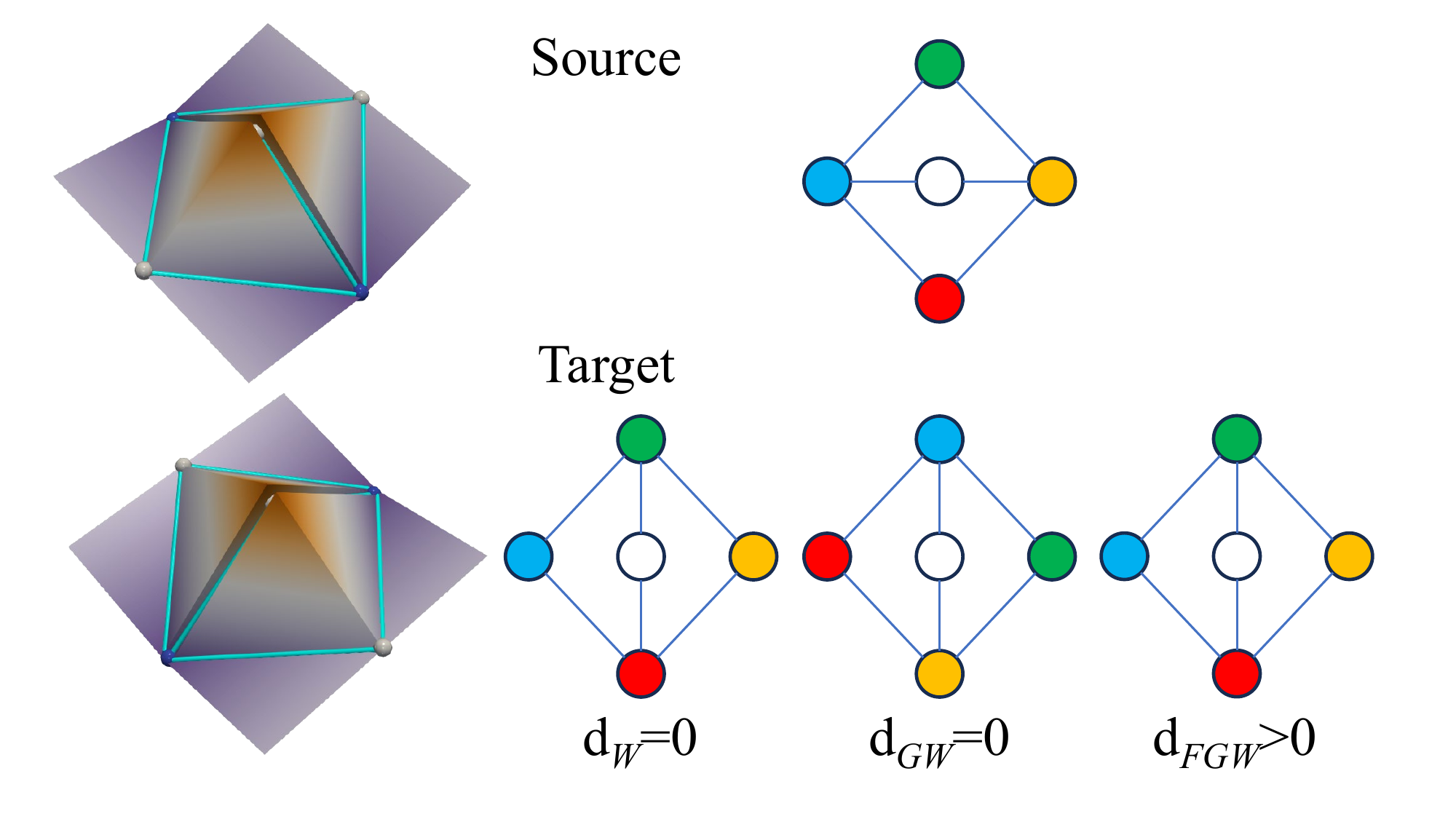}
    \vspace{-2mm}
    \caption{Left: synthetic scalar fields, together with their corresponding Morse graphs (in cyan). 
    Right: Morse graphs of the scalar fields; colormap shows structural correspondences under various OT-type metrics.}
    \label{fig:ot-example}
    \vspace{-2mm}
\end{figure}

We use an example in~\cref{fig:ot-example} to explain the intuition behind the Wasserstein distance, GW distance, and FGW distance. 
We consider two scalar fields shown on the left,  
with corresponding Morse graphs shown on the right. 
To compute OT-type distances, we use the geometric locations of nodes as attributes, and the shortest geodesics between nodes to describe the network structures; see Sec. 4 for details.

We observe the behavior of three OT-type distances based on the node color correspondences. 
The Wasserstein distance is $0$ because the locations of matched nodes in the source and target are identical. 
The GW distance is also $0$ since it finds a coupling such that the graph structure in the target is identical to the source. 
Such a coupling rotates the node color correspondence by $90^{\circ}$.
However, since the FGW distance takes both graph structures and geometric locations into consideration, it becomes non-zero in this example. 


%% file: sec-datasets.tex
\section{Details on Datasets}
\label{sec:datasets}

The synthetic {\SN} dataset is formed by the sum of two sine waves. 
The target field is generated with additional noise in comparison with the source field. 

The {\WD} dataset includes $15$ wind velocity fields from the IRI/LDEO Climate Data Library: \textit{pressure\_level\_wind} is obtained via the NCEP CFSv2 Ensemble~\cite{NCEPData} with forecasted and perturbed parameters. 
It focuses on a spatial range of $150^\circ$W-$49.5^\circ$W and $90^\circ$N-$10^\circ$S at a pressure level $200$ hPA and a forecast hour $0$ on January 01, 2015. 

The {\HC} dataset~\cite{cgl} comes from a simulation via the Gerris flow solver~\cite{gerrisflowsolver}. 
The simulation shows a 2D flow of a heated cylinder using the Boussinesq approximation~\cite{GuntherGrossTheisel2017}. 
It describes a time-varying turbulent plume containing multiple small vortices. 
We select $100$ time steps from the original dataset ($800$-$899$ from $2000$ time steps) and compute the Morse graphs from the velocity magnitude fields.

The {\NS} dataset is a direct numerical Navier Stokes simulation available from~\cite{iCFDDatabase}. 
Camarri \etal~\cite{camarri05} used a version of the simulation that Tino Weinkauf had uniformly resampled,  which was used for smoke visualization by von Funck \etal~\cite{vonfunck08a}. 
The dataset is a 3D time-varying fluid flow simulation  around a square cylinder placed symmetrically between two parallel walls. 
We select time steps $60$-$66$ and a 2D slice perpendicular to the z-axis ($z=24$) and compute Morse graphs from the velocity magnitude fields. 

The {\RS} dataset comes from the IEEE 2020 SciVis Contest~\cite{sciVis2020} and shows an eddy simulation of the Red Sea.  
It uses the MIT ocean general circulation model (MITgcm) and the Data Research Testbed (DART)~\cite{HOTEIT20131} to create an ensemble with varying initial conditions. 
A 3D area with a resolution of 500$\times$500$\times$50 is sampled across 60 time steps~\cite{toye2017ensemble}. 
We use 10 ensemble members from 2D slices perpendicular to the z-axis ($z=1$) at time step 40. 
We compute the Morse graphs from the velocity magnitude fields.

%% file: sec-parameters.tex
\section{Preprocessing and Parameter Setting}
\label{sec:parameters}

\para{Persistence simplification.}
We apply \emph{persistence simplification} to each dataset before computing the Morse graphs.
We normalize the range of a given scalar field to be $[0,1]$ and use $\varepsilon \in [0,1]$ to denote the simplification threshold.  
$\varepsilon$ is chosen based on the \emph{persistence graph}~\cite{GerberBremerPascucci2010}, where the x-axis represents $\varepsilon$, the y-axis captures the number of local maxima (in our setting), and a plateau implies a stable range of scales to separate features from noise. 
An $\varepsilon$-simplification means that the scalar field is simplified such that critical points with persistence less than $\varepsilon$ are removed.  

\begin{figure}[h]
    \centering
    \includegraphics[width=1.0\columnwidth]{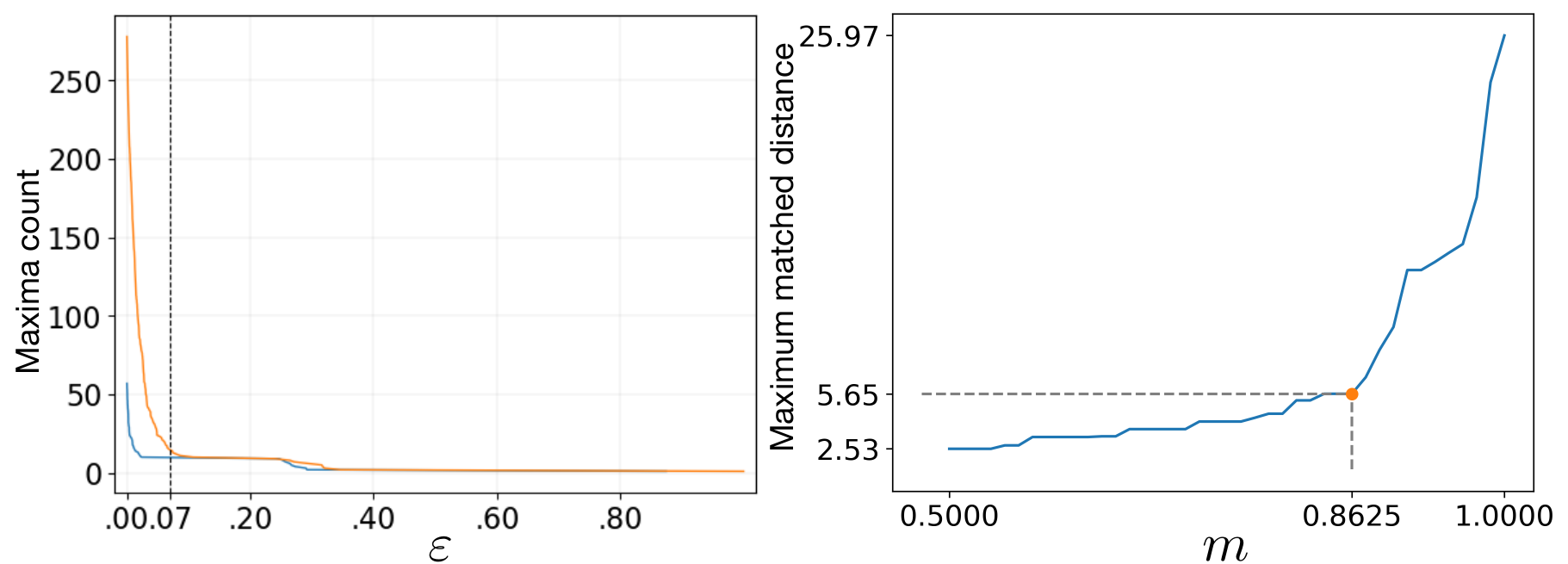}
    \vspace{-6mm}
    \caption{Parameter tuning for the {\SN} dataset. Left: persistence simplification using a persistence graph, where $\varepsilon=7\%$. Right: $m=0.8625$ is the elbow  point based on the maximum matched distances.}
    \label{fig:parameter-tuning}
    \vspace{-2mm}
\end{figure}

For the {\SN} dataset, \cref{fig:parameter-tuning} (left) shows the simplification threshold $\varepsilon = 7\%$ chosen at a plateau for both the source field (blue) and the target field (orange). 
The parameter $\varepsilon$ is chosen to be $3\%, 10\%, 7\%$, and $1\%$ for the \WD, \HC, \NS, and {\RS} datasets, respectively. 

\para{Parameter setting for FGW distance.}
In this paper, the attribute space is set to be $\Rspace^2$ endowed with a Euclidean distance, and the attribute function assigns a node its location in the domain. 
We apply normalization such that the Euclidean distance between a node in the source Morse graph and a node in the target Morse graph falls within $[0,1]$. 
We also rescale elements in each network function matrix $W$ such that they fall into the range $[0, 1]$. 
With such a normalization, we set $\alpha=0.5$ when computing the FGW and pFGW distances, striking a balance between the Wasserstein component and the GW component.

\para{Node probability distribution.}
We assign a probability distribution to all nodes in the Morse complexes when we model them as measure networks. 
Without any prior knowledge about the space, a uniform probability distribution is a reasonable option for computing similarities between spaces based on previous works (e.g.~\cite{Memoli2011,ChowdhuryNeedham2020,demetci2022scot,LiYanYan2023}), since it ensures that all nodes in the space have equal importance in the distance. 


\begin{figure}[!ht]
    \centering
    \includegraphics[width=0.95\columnwidth]{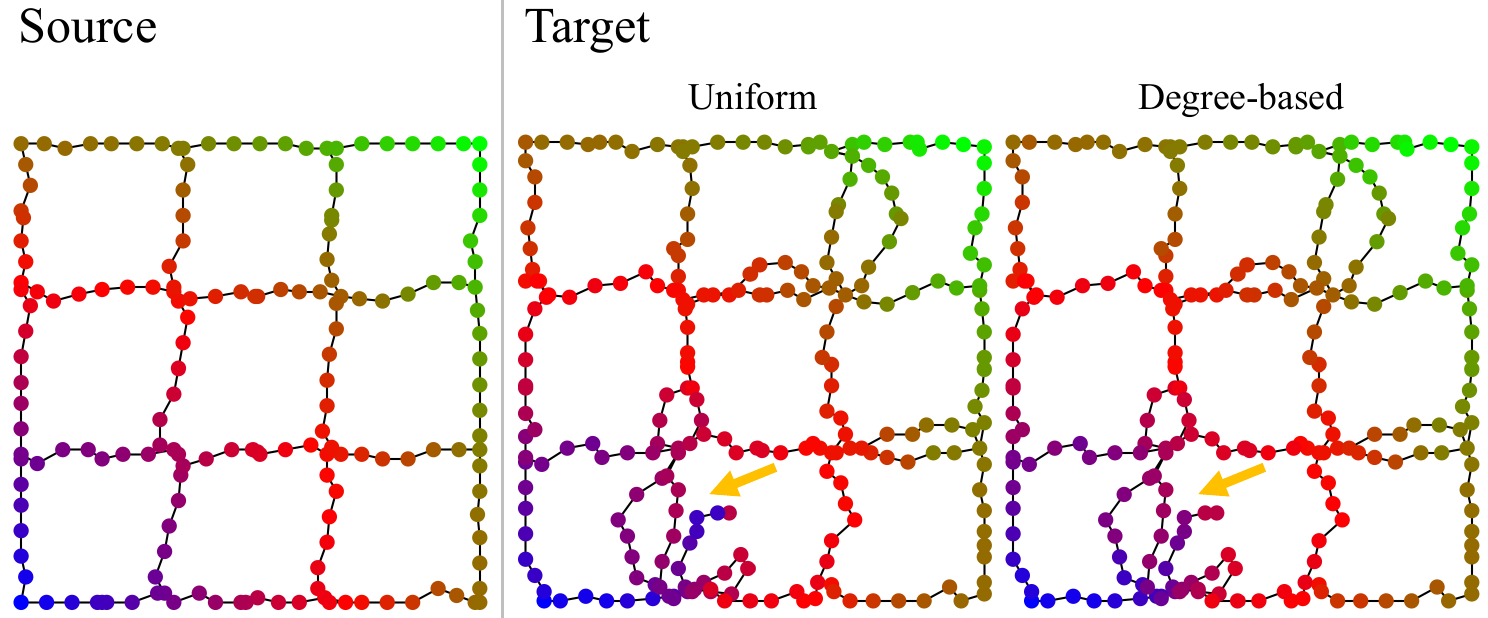}
    \vspace{-2mm}
    \caption{FGW distance for the {\SN} dataset with different node probability distributions.}
    \label{fig:node-prob-distribution}
    \vspace{-2mm}
\end{figure}

In this paper, we apply a probability distribution that is  proportional to the degree of nodes instead of the uniform distribution. 
If we apply a uniform distribution, due to our node sampling strategy, the number of nodes along each edge may be sensitive to minor changes in the geometry of an edge, thus affecting the distribution.  
Furthermore, newly appeared or disappeared edges along with their sampled nodes may also affect the probability of existing nodes. 
By applying a degree-based probability distribution, nodes with higher degrees (typically critical points)  serve as anchor points that are considered more important than the (sampled) regular points; this approach has also been prevalent in the graph OT literature, due to its superior empirical performance for certain tasks (e.g., \cite{XuLuoCarin2019,chowdhury2021generalized}).
For example, in~\cref{fig:node-prob-distribution}, the result using the FGW distance with uniform distribution shows undesired matching as indicated by the orange arrow. 
The degree-based distribution improves the coupling in the same area, showing a smoother color trend.

\begin{figure}[h]
    \centering
    \includegraphics[width=0.95\columnwidth]{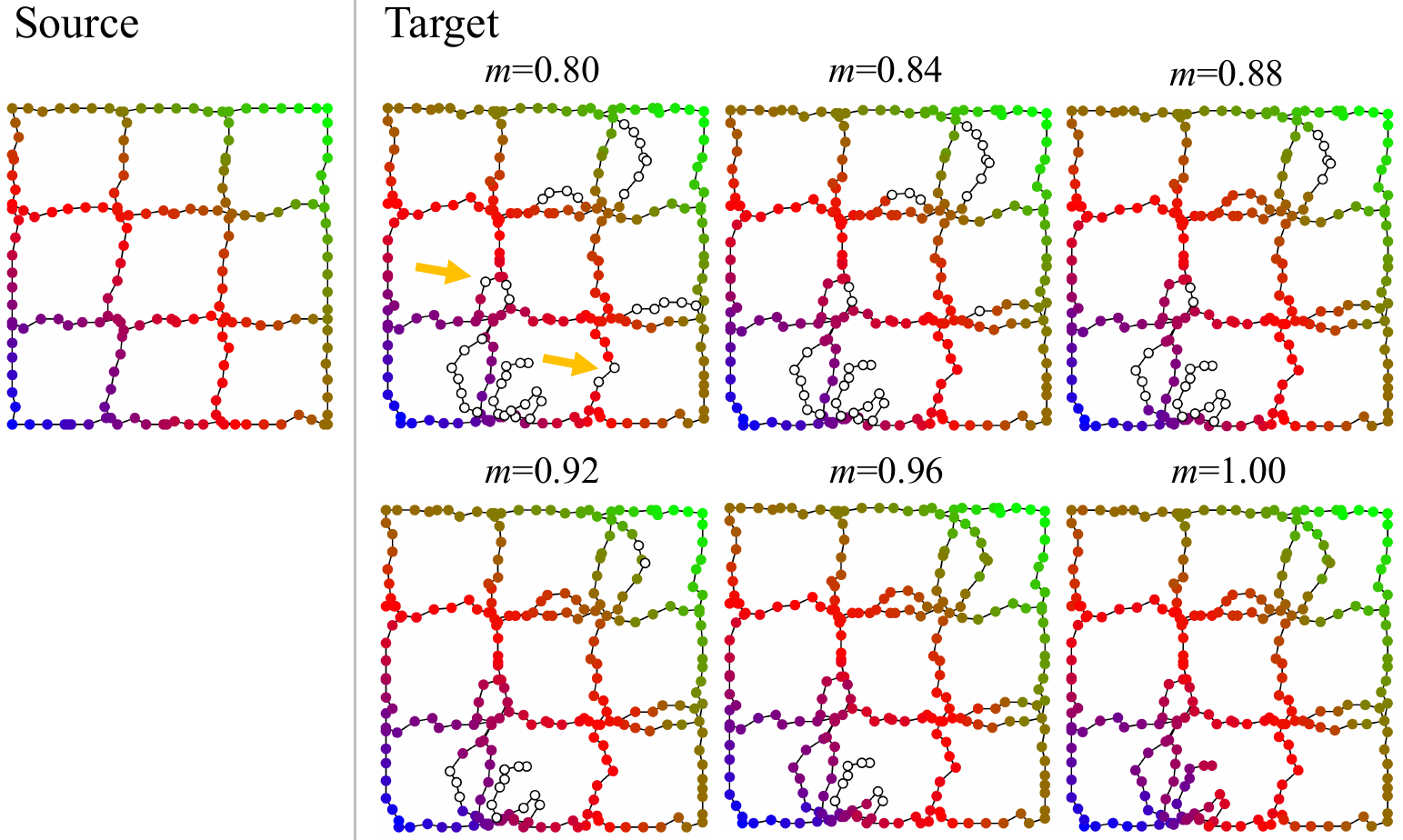}
    \vspace{-2mm}
    \caption{pFGW distance for the {\SN} dataset with different $m$ values.}
    \label{fig:partial-parameter}
    \vspace{-2mm}
\end{figure}

\para{Partial OT parameter tuning.}
We explain the parameter tuning of partial OT via the {\SN} example in~\cref{fig:partial-parameter}, following a previous work~\cite{LiYanYan2023}. 
As the parameter $m$ decreases from $1.00$ to $0.80$, the number of hollow nodes in the target graph (e.g., nodes that are ignored during the coupling) increases gradually. 
During this process, nodes in the target graph on the noisy edges are ignored first, as matching these nodes to any node in the source graph leads to a large distance. 
This is desirable as we apply partial OT to ignore noisy features.

However, when $m$ reaches $0.80$, important nodes (as indicated by orange arrows) start to be ignored. We aim to keep the important nodes in the coupling because they describe the main structure of the target graph.
Therefore, our high-level idea in tuning $m$ is to strike  a balance between ignoring noise and preserving main features. 
In other words, we would like to maintain as much mass as possible in the coupling result while removing unreasonable matchings.

We assume that nodes are matched based on their Euclidean proximity. 
That is, matching nodes that are far apart is undesirable. 
Using partial OT, nodes that are matched faraway from each other are ignored first. 
By ignoring these nodes, the maximum Euclidean distance between the matched nodes in the source and the target drops drastically. 
Therefore, we plot the maximum Euclidean distance between the matched points w.r.t. $m$, referred to as the \textit{maximum matched distance}. 
We use the ``elbow'' method to select an $m$ where the maximum matched distance is not too high. 
As shown in~\cref{fig:parameter-tuning} (right), for the {\SN} dataset, such a distance grows significantly after the elbow point at $m=0.8625$. 

We understand that such a strategy comes with some limitations. 
There may be multiple ``elbows'' in the plot, and a unified $m$ may not work well for all pairs of Morse graphs. 
Without prior knowledge about the data, the choice of $m$ may not be unique. Different $m$ values indicate different tolerance for noises, which may all be meaningful. 

%% file: sec-runtime.tex
\section{Runtime Analysis}
We report the runtime in computing the OT-type distances between the source Morse graph and the target Morse graph for all real-world datasets in~\cref{table:runtime}. 
All these distances are easy and efficient to compute. 
Relatively speaking, for a fixed dataset, the GW distance takes the most time. 
The runtime was collected on an Arch Linux system with an Intel(R) Core(TM) i7-6700K 4.00 GHz CPU with 32 GB memory.

\begin{table}[!ht]
\begin{center}
\vspace{-2mm}
\resizebox{1.0\columnwidth}{!}{
\begin{tabular}{c|c|cccccc}
\hline
\textbf{Dataset} & \multicolumn{1}{l|}{\textbf{\# of nodes}} & \textbf{W} & \textbf{GW} & \textbf{FGW} & \textbf{pW} & \multicolumn{1}{l}{\textbf{pGW}} & \multicolumn{1}{l}{\textbf{pFGW}} \\ \hline
Wind             & 99                                        & 0.081                & 0.348       & 0.118        & 0.079       & 0.129                            & 0.098                             \\
Heated Cylinder  & 601                                       & 1.662                & 10.146      & 6.425        & 1.451       & 1.776                            & 5.123                             \\
Navier-Stokes    & 130                                       & 0.184                & 0.278       & 0.130        & 0.118       & 0.146                            & 0.178                             \\
Red Sea          & 171                                       & 0.183                & 0.531       & 0.250        & 0.166       & 0.283                            & 0.286                             \\ \hline
\end{tabular}
}
\end{center}
\vspace{-6mm}
\caption{Runtime (in seconds) for OT-type distances between the source and the target Morse graphs across all real-world datasets. }
\label{table:runtime}
\vspace{-6mm}
\end{table}

%% file: sec-temporal-analysis.tex
\begin{figure*}[!ht]
    \centering
    \includegraphics[width=0.9\textwidth]{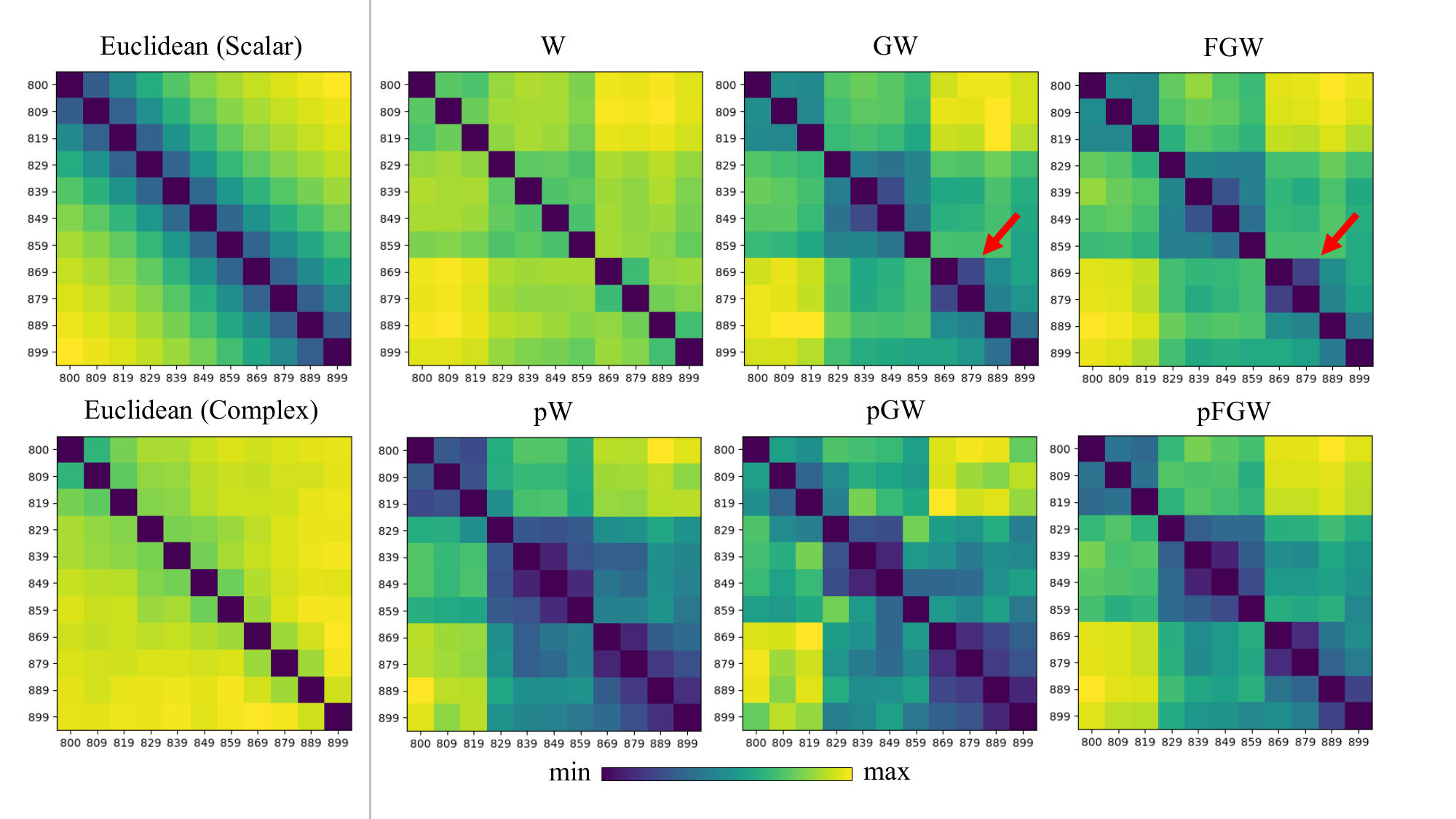}
    \vspace{-2mm}
    \caption{Pairwise distance matrix for two Euclidean metrics and six OT-type distances. Labels for both axes of matrices are time steps.}
    \label{fig:hc-temporal-matrix}
    \vspace{-6mm}
\end{figure*} 

\begin{figure}[!ht]
    \centering
    \includegraphics[width=0.9\columnwidth]{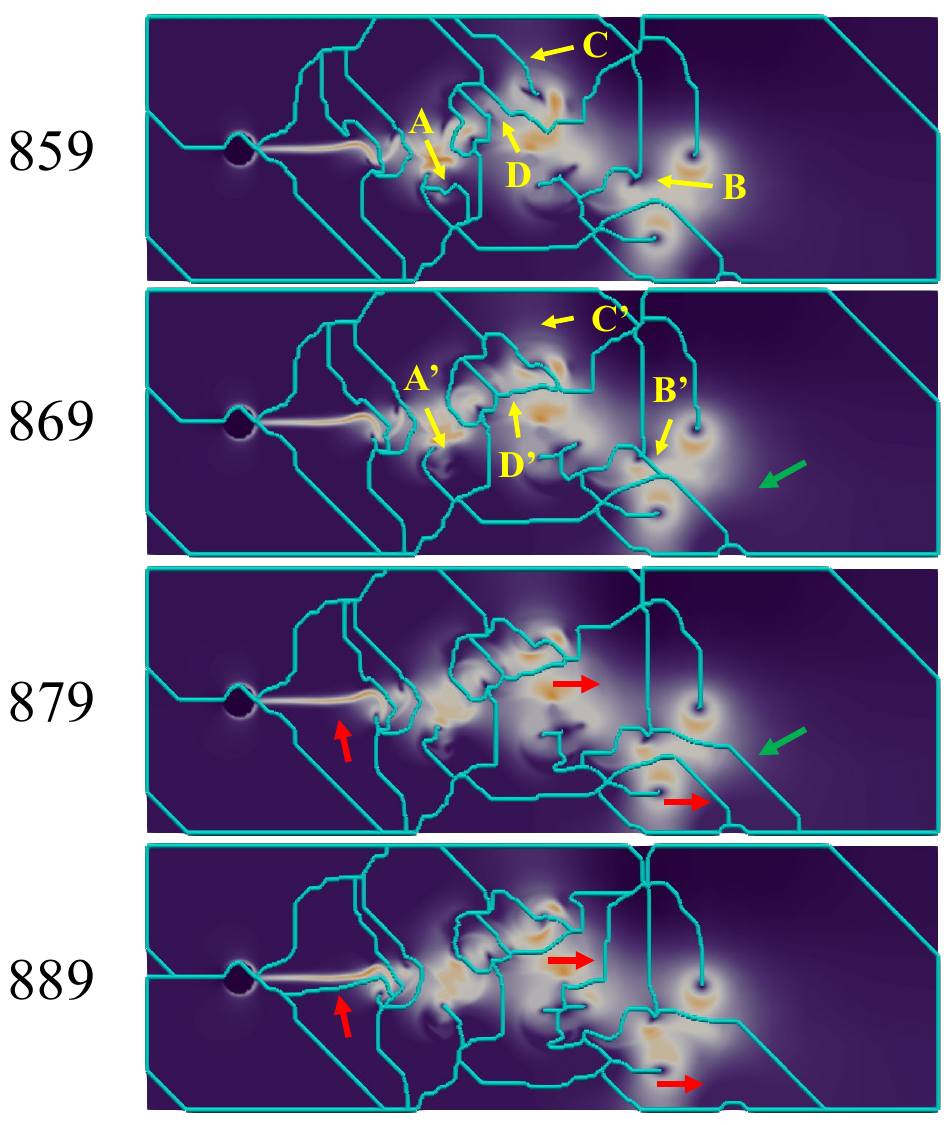}
    \vspace{-2mm}
    \caption{Scalar fields and Morse complexes in the {\HC} dataset at time step $859$, $869$, $879$, and $889$.}
    \label{fig:hc-temporal-screenshot}
    \vspace{-2mm}
\end{figure}

\section{Comparison with Euclidean Metrics}
\label{sec: temporal-analysis}

OT-type distances can be applied to various visualization tasks. In this section, we use the {\HC} dataset to demonstrate its application in detecting topological (\ie, Morse complex) changes in a time-varying scalar field. 
Since the topological variation between adjacent time steps is quite small, we use scalar fields at time step $800$, $809$, $819$, $829$, $839$, $849$, $859$, $869$, $879$, $889$ and $899$ in the analysis. 


There are few existing methods suitable for comparing Morse complexes. 
We include two simple Euclidean metrics as baseline methods for comparison. 
Suppose a scalar field $f$ is uniformly sampled on a regular grid and represented as a matrix. 
The first metric is the Frobenius norm between the matrix representations of scalar fields, referred to as the \emph{Euclidean scalar distance}. 
For the second metric, we map a Morse complex onto a binary matrix. 
An entry (that corresponds to a grid cell) in the matrix is $1$ if it overlaps with the Morse graph; otherwise, it is $0$.  
We then compute the Frobenius norm between the binary matrix of Morse complexes as the second  metric, referred to as the \emph{Euclidean complex distance}.

\cref{fig:hc-temporal-matrix} shows the pairwise distance matrix of each distance. 
Two matrices on the left are from the two Euclidean metrics.
Not surprisingly, we cannot detect topological changes using these Euclidean metrics. 

On there other hand, all OT-type distance matrices in~\cref{fig:hc-temporal-matrix} (right) show patterns indicating topological changes. Among the three non-partial distances, both GW and FGW distances show clear block structures (indicated by red arrows) in the matrix, indicating similarity in topology.

We further explore the data within such a block structure.~\cref{fig:hc-temporal-screenshot} shows the scalar fields along with Morse complexes from time step $859 \to 889$. 
Changes in Morse complex at $859 \to 869$ are highlighted by yellow arrows. 
The edge $A$ at time step $859$ disappears at time step $869$ (as indicated by $A'$). 
In the area indicated by $B$ at time step $859$, an edge appears as $B'$ at time step $869$. 
Similarly, $C \to C'$ indicates the disappearance of a long edge, whereas $D \to D'$ indicates a new loop enclosed by a new edge.
Such structural changes are captured by the GW and FGW distances, resulting in high distances. 
Similarly, significant structural changes (\ie, edge appearances and disappearances) at $869 \to 879$ and at $879 \to 889$ are highlighted by green and red arrows, respectively.
We observe only one green arrow at $869 \to 879$, whereas we see four yellow arrows at $859 \to 869$ and three red arrows at $879 \to 889$. 
This observation is consistent with the result of GW distance and FGW distance. 
In comparison, among four time steps from $859$ to $889$, the Morse complexes between time step $869$ and $879$ are relatively similar, whereas time step $889$ is more different, time step $859$ the most different. 



%% file: sec-partial-definition.tex
\section{Definition of Partial OT-type Distances}
\label{sec-partial-definition}

Following Sec. 3, we provide the formal definition of partial OT-type distances.

\para{Partial Wasserstein distance \cite{ChapelAlayaGasso2020}.} Given $m \in [0,1]$, the \emph{$q$-partial Wasserstein distance} (pW) is defined as 
\begin{align}
d^{pW}_{q,m}(G_{1},G_{2})^q = \min_{C\in\Ccal_m} \sum_{i,j} d_A(a_i,b_j)^q C_{i,j}.
\label{eq:pw-distance}
\end{align}
\para{Partial Gromov-Wasserstein distance \cite{ChapelAlayaGasso2020}.}
The \emph{$q$-partial Gromov-Wasserstein distance} (pGW) is   
\begin{align}
d_{q,m}^{pGW}(G_1, G_2)^q =  \min_{C\in\Ccal_m}\sum_{i, j, k, l}|W_1(i, k) - W_2(j, l)|^{q} C_{i,j}C_{k,l}.
\label{eq:pgw}
\end{align}	
\para{Partial Fused Gromov-Wasserstein distance \cite{LiYanYan2023}.} 
For parameters $\alpha,m \in [0,1]$, the \emph{$q$-partial Fused Gromov-Wasserstein distance} (pFGW) is
\begin{align}
& d^{pFGW}_{q,\alpha,m}(G_{1},G_{2})^q = \min_{C\in\Ccal_m}  \sum_{ i,j,k,l} [(1-\alpha) d_A(a_i,b_j)^q \nonumber \\
&\qquad \qquad\qquad \qquad + \alpha |W_1(i,k) - W_{2}(j,l))|^q] C_{i,j}C_{k,l}.
\label{eq:pfgw}
\end{align}
Choosing parameters appropriately, we see that all distances defined previously can be seen as special cases of the pFGW distance. 

%% file: sec-metric-property.tex
\section{Metric Properties of OT-type Distances}
\label{sec-metric-property}

The Wasserstein and GW distances are pseudometrics on the space of $A$-attributed measure networks (i.e., they are symmetric, satisfy the triangle inequality, and vanish when $G_1 = G_2$), see~\cite{chowdhury2019gromov}. 
For $\alpha \in (0,1)$, the FGW distance is a pseudometric when $q=1$. 
It is symmetric, satisfies $d^{FGW}_q(G,G) = 0$, and meets a modified triangle inequality when $q \geq 2$:
\begin{align}
d^{FGW}_q(G_1,G_3) \leq 2^{1-\frac{1}{q}} (d^{FGW}_q(G_1,G_2) + d^{FGW}_q(G_2,G_3)).
\label{eq:fgw-property}
\end{align} 
Indeed, this follows from \cite[Theorem 1]{vayer2020fused} (this is proved assuming that the $W$-functions are metrics, but the proof goes through without this assumption). 
The partial versions of the distances, however, fail to satisfy the triangle inequality (or any simple variant) when $m \neq 1$. 